# Opinion on sustainability-performance relationship: Na-ion batteries versus Li-ion batteries

Siripak Sangsinsorn,[a] Sonia Dsoke[a,b] and Oana Cojocaru-Mirédin*[a]

[a] S. Sangsinsorn, Prof. S. Dsoke, Prof. O. Cojocaru-Mirédin
University of Freiburg, Department of Sustainable Systems Engineering (INATECH), Emmy-Noether-Strasse 2, 79110 Freiburg im Breisgau
[b] Prof. S. Dsoke
Fraunhofer Institute for Solar Energy Systems, Department of Electrical Energy Storage, Heidenhofstr. 2, 79110 Freiburg, Germany.

Corresponding author: oana.cojocaru-miredin@inatech.uni-freiburg.de

**Abstract:** A new era for energy storage devices, such as rechargeable batteries, has been opened in the last decades. However, commercially available energy storage devices are based mainly on critical elements such as Li, Co, Mn, P, Ni, and graphite opening sustainability concerns for the industry and society. Yet, these elements are crucial for both, battery cells (energy density, voltage and cycle lifetime) as well as for electrode and electrolyte material properties (specific capacity, working potential, ionic and electronic conductivity). New research directions point toward the development of materials free of such critical elements, but unfortunately often at the price of lower performance. In the present work, we study the interlink between sustainability and properties at both, material and battery cell levels of Li-ion batteries (LIBs) and Na-ion batteries (SIBs), aiming to address the research gap in simultaneously comparing sustainability and performance. Sustainability factors, such as supply and environmental risks, are assessed based on the critical element content in the cells and the individual risks associated with these materials. Although LIB cells exhibit superior energy density compared to their Na-ion counterparts, their sustainability is compromised by the consumption of Li and graphite. SIBs do not work with graphite like LIBs due to the difficulty of intercalating Na within the structure of graphite. They work with other carbonaceous materials (i.e. hard carbon) that can be produced in a sustainable way from biomass. Moreover, the composition of cathode materials significantly influences the sustainability of SIBs cells, with cells containing P and V presenting notable supply and environmental concerns. SIBs, with cathodes such as a layered typed Na-metal oxide which contains Ni and Mn as redox centers (Na$_x$(Ni,Mn)O$_2$) such as Na$_{1.1}$(Ni$_{0.3}$Mn$_{0.5}$Mg$_{0.05}$Ti$_{0.05}$)O$_2$ (NMMT) and Na$_2$FeFe(CN)$_6$ (NPBA) and anodes such as hard carbon, demonstrate exceptional sustainability scores. Their potential energy densities are projected to match that of a LiFePO$_4$ (LFP)-based LIB, making them increasingly attractive for commercial applications.

## 1. Introduction

Batteries are key players in achieving net-zero CO$_2$ emissions, particularly through energy transition and electrification measures. However, the demand for batteries necessitates a significant amount of minerals, leading to sustainability concerns. While numerous studies focus on comparing either the sustainability or the performance of certain battery chemistries, research works that evaluate both aspects simultaneously are scarce. This study aims to bridge this gap by evaluating and comparing the sustainability aspects and performance of LIBs, the most commonly used type, with those of emerging batteries in the research field such as Na-ion batteries (SIBs).

The material-sustainability encompasses two dimensions: economic, and environmental impacts. When discussing sustainability, the raw materials extracted from minerals are in focus because the minerals can only be found in specific areas or countries and mining is a critical process that negatively impacts the environment. This is particularly true for high raw material content batteries.[1,2] These raw materials extracted from minerals are limitedly available in some regions leading to economic concerns such as a high supply risk recognized as the primary economic indicator in this context.[3,4] Additionally, global mining potentially affects approximately 50 million square kilometers of the Earth's surface and poses threats to biodiversity.[1]

Researches highlight that LIBs pose supply risks because they contain critical elements such as Co, Li, P, Ni and graphite.[5,6] Therefore, the community has developed alternative Li-free batteries such as SIBs[7] given that Na is the sixth most abundant material in the Earth's crust[8] and available in seawater and offers a similar electrochemical working principle as Li in rechargeable batteries.[9] Additionally, unlike LIBs, SIBs do not function with graphite as anodes but instead with hard carbon. Thankfully, the hard carbon can be synthesized from many biowaste sources and does not need minerals as its precursors, significantly reducing the supply risks.[10] Yet, considering fixing the negative electrode with the hard carbon, the positive electrode material should be taken into consideration and evaluated in terms of environmental impact. For instance, when phosphate-type cathodes, such as Na$_3$V$_2$(PO$_4$)$_3$ (NVP) for SIBs and LFP for LIBs, are compared, the NVP cathode poses a higher environmental risk compared to the LFP cathodes.[11] In this analysis, we do not consider the same analog NaFePO$_4$ (NFP) due to its current lack of commercial viability. This is primarily because of its synthesis challenges and commercialization barriers, including low thermal stability, low intrinsic electronic conductivity, and relatively large volume expansion.[12]

Besides these environmental concerns, the device and materials properties are equally essential to meet the technological requirements. For example, the specific energy of batteries directly impacts the operational costs and environmental footprint of electric vehicles.[13,14] It is without any doubt that currently the highest specific energy is reached for the LIBs.[13,14] While SIBs have a specific energy of approximately 160 Wh/kg,[15,16] they are still competitive with LFP batteries for, which the specific energy ranges from 90 to 160 Wh/kg.[17]

Moreover, factors such as cycle life and energy retention are vital for longevity. Longer battery life leads to reduced material use, decreased eco-toxicity, and lower energy demands for new battery production.[15,16] The life cycle of SIBs ranges between 4,000 to 7,000 cycles, similar to that of LIBs.[11]

Last but not least, ionic and electronic conductivities are key material properties for the battery cycling performance (rapid charging and discharging rates); properties essential not only for liquid or solid electrolytes but also for cathode materials.[18] It is expected indeed theoretically that Na ionic conductivity in these battery materials is very high due to their non-compact crystal structures (such as layered, NASICON or spinel-type structure[19]). This has been proven for example for Na-ion liquid electrolytes, which exhibit higher ion conductivity values than that of the Li counterpart.[20] Yet, it has been demonstrated that the Na ionic conductivity in solid electrolytes and various cathode materials is low, much lower than corresponding Li-based compounds.[20,21,22,23–30]

Hence, in this work, we are going to explore the environmental parameters for the existing Li-ion and Na-ion-based materials and devices and corroborate them with the corresponding device and materials properties. For that, first, numerical equations used for the description of the supply and environmental risks will be provided. Second, the existing device and material properties will be described. Finally, the correlation between sustainability (supply and environmental risks) and the properties of battery cells (energy density, voltage, and lifetime) as well as of materials (potential window and ionic conductivity) will be made.

## 2. Methodology

### 2.1 Environmental parameters

#### a) Supply risk (SR) of a raw material

Despite the existence of various methods to assess supply risk, which incorporate different criteria[3,4,6,31,32], there is a consensus in their interpretations: a higher supply risk signifies a greater criticality of the material and the opposite. The U.S. Department of Energy and the European Commission have developed distinct methodologies for evaluating regional supply risks of critical materials.[31–33] The U.S. Department of Energy utilizes a weight-averaged method, factoring in basic availability, demand from competing technologies, political, regulatory, and social influences, dependencies on other markets, and producer diversity.[32] Risk levels are categorized on a scale from 1 to 5, with level 5 representing the highest risk. Conversely, the European Commission employs a multiplicative approach, aggregating effects of geopolitical supply risks, trade parameters, import reliance, end-of-life recycling input rates, and substitution indices.[31,33] A material is considered critical when its supply risk score exceeds a threshold of 1.

This study evaluates the supply risk of materials ($SR_m$) using data from the Study on the Critical Raw Materials for the EU 2023 report[33], and follows the methodology provided in European Commission's guidelines as depicted in Equation 1[31]:

$$SR_m = \left(HHI_{WGI,t_{GS}} \times \frac{IR}{2} + HHI_{WGI,t_{EU}}\left(1 - \frac{IR}{2}\right)\right) \times (1 - EoL_{RIR}) \times SI_{SR} \quad (1)$$

The primary assessment focuses on country concentration risks, evaluated via the Herfindahl-Hirschman Index (HHI).[31,34] This index gauges risk by considering the market shares of countries ($S_i$ for global supply (GS) and $S_j$ for EU-specific supply (EU)) alongside their governance indicators ($WGI_I$ and $WGI_j$, respectively), as determined by the World Bank. The formula also accounts for the impact of trade-related variable ($t_I$ or $t_j$) of raw materials destined for Europe, specifically tailored for the European Union context, as represented in Equations 2[31]:

$$HHI_{WGI,t_{GS}} = \sum (S_i)^2 \times WGI_i \times t_i$$
$$HHI_{WGI,t_{EU}} = \sum (S_j)^2 \times WGI_j \times t_j \quad (2)$$

Equation 3[31] describes the origin of trade-related variable ($t_I$ or $t_i$) which can be derived from accounts for export taxes ($ET - T_{ac}$), from export prohibitions ($EQ_c$), from export quotas ($EP_c$), and from a base value when the supplier is an EU member ($EU_c$).

$$t_{i,j} = ET - T_{ac} \text{ or } EQ_c \text{ or } EP_c \text{ or } EU_c \quad (3)$$

The import reliance (IR) measures the dependency on imports for a particular material, comparing the net imports to the apparent consumption as shown in Equation 4[31]:

$$IR = \frac{Import - Export}{Domestic\ production + Import - Export} \quad (4)$$

In Equation 5[31], the substitution index ($SI_{SR}$) evaluates the potential to replace a material with alternatives, factoring in the sufficiency of global production ($SP_i$), the criticality of the alternative materials ($SCr_i$), and whether the substitution is a primary or co/by-product ($SCo_i$)

$$SI_{SR} = \sum (SP_i \times SCr_i \times SCo_i)^{\frac{1}{3}} \times \sum (Sub - Share_{i,a} \times Share_a) \quad (5)$$

The last factor is the end-of-life recycled input rate ($EoL_{RIR}$) presenting the ratio between the recycled material used as the secondary input to EU ($Input_{sec}$) and the demand for the material in EU including the primary input ($Input_{pri}$) and secondary input to EU[31]:

$$EoL_{RIR} = \frac{Input_{sec}}{Input_{pri} + Input_{sec}} \quad (6)$$

#### b) Aggregated supply risk of a battery cell

The supply risk at the cell level is calculated by aggregating the risks of all elements based on their mass share within the cell, a method referred to as the mass share approach, as shown in Equation 7[4,35] below:

$$SR = \sum_{m=1}^{M} SR_m \times \%wt_m \qquad (7)$$

Where $\%wt_m$ represents the weight percentage of the material m which is interchangeable with mass percentage because the gravity constant cancels out in this calculation.

**c) Environmental risk of a raw material**

The environmental risk of raw materials ($ER_m$) is conducted according to European Commission[36] as given in Equation 8 below, given that the environmental performance index is accredited, acceptable, and utilized in national official reports[37]. In addition, the database for the calculation of this value is the same as the database of the calculation of the supply risk (SR) allowing a reliable comparison of these two values. The $HHI_{EPI_{GS}}$ is an indicator of market concentration from global suppliers, considering both their market share and environmental performance index (EPI). Similar to the supply risk calculation, indices for end-of-life and substitution are also included to evaluate the necessary material amounts.

$$ER_m = HHI_{EPI_{GS}} \times (1 - EoL_{RIR}) \times SI_{SR} \qquad (8)$$

**d) Aggregated environmental risk of a battery cell**

The method for the calculation of the aggregated environmental risk is based on mass weighting method; similar to the aggregated supply risk shown in Equation 9 below:

$$ER = \sum_{m=1}^{M} ER_m \times \%wt_m \qquad (9)$$

The evaluation focuses on potentially commercial cells, including Li-ion and Na-ion types. For simplicity, the cell are named based their cathode material as the anode consistently carbon-based (graphite for Li and HC for Na battery cells). While some SIBs use PBA as both cathode and anode materials (e.g., Na battery cells proposed by Natron Energy), they are not within the scope of this study. Although we initially considered studying a commercial layered transition metal cathode for SIBs, such as $Na_{0.677}(Ni_{0.3}Mn_{0.6}Mg_{0.033}Ti_{0.067})O_2$ [38], there is insufficient data to accurately assess their supply and environmental risks. Consequently, we select a similar compound $Na_{1.1}(Ni_{0.3}Mn_{0.5}Mg_{0.05}Ti_{0.05})O_2$ (NMMT) which shares the same $Na(TM)O_2$ structure where "TM" denotes an individual or group of transition metals, for evaluation.

In sum, all cells used in this study, including $LiNi_{0.8}Mn_{0.1}Co_{0.1}O_2$ (NMC811), $LiFePO_4$ (LFP), $Na_{1.1}(Ni_{0.3}Mn_{0.5}Mg_{0.05}Ti_{0.05})O_2$ (NMMT), $Na_3V_2(PO_4)_2F_3$ (NVPF), $Na_2FeFe(CN)_6$ (NPBA), $Na_4MnV(PO_4)_3$ (NMVP) are of the same geometry namely 21700 cylindrical cells. The key parameters are shown in Table 1.

**2.2 Properties of cells and their respective materials**

A desirable battery cell should have high specific energy and a long life cycle. Specific energy is a key metric for evaluating its performance, particularly critical in the mobility sector, where applications demand high energy storage capabilities within a lightweight framework.[39] Moreover, a longer cycle lifetime of battery cells has direct implications on sustainability; i.e. a longer cycle lifetime reduces the need for new production and, thus, minimizes the consumption of mineral resources. A cell consists of anode, cathode, electrolyte, separator, current collectors, and housing. The components rely on raw materials extracted from minerals, whose supply and environmental risks are evaluated, while the separator, synthesized from polymers, is outside the scope of this study. Table 1 provides significant information for evaluating the battery cells, including materials, mass, specific energy, voltage, cycle lifetime, and potential toxicity connected to the production process.

The mass of the total cells and their components is taken from the literature[40] calculated based on cell dimensions, mass loading, and the physical properties (density, porosity) of materials. For cathode-active materials containing several elements, the mass of an element ($m_i$) in the cathode active material is assumed to be propotional to its mole fraction in the compound, based on the chemical formula of the compound. Thus, it can be calculated using Equation 10.

$$m_i = m_{CAM} \times \frac{x_i M_i}{\sum_i x_i M_i} \qquad (10)$$

Here, $m_{cam}$ represents the mass of cathode active materials (excluding the current collector), $x_i$ is the mole fraction of element i, and $M_i$ is the molar mass of element i.

Although the literature provides the specific energy calculated from the specific capacity of the cathodes (given the voltage at 50% state of charge (SOC), depth of discharge, round trip efficiency, and total cell mass), in this study the specific capacity and voltage of the cathodes are updated based on the other reference[41], providing the values within the ranges in our database, see Table 2.

NMC811 offers the highest specific energy and voltage among the sample cells. However, its cycle lifetime of 4,000 cycles is lower than that of LFP and Na-ion battery cells, including NVPF, NMVP, and NPBA.[41] SIBs potentially reach the current LFP specific energy.[40] For instance, NMMT and NPBA have higher predicted specific capacities than LFP in the current state.[42] NPBA, in particular, is more attractive since it has a higher cycle life than that of NMMT.[41] In addition, the anode of Li and Na-ion battery cells is made from graphite and hard carbon, respectively.[41] The natural graphite is from mining whereas the hard carbon is chemically synthesized from bio sources, canceling the risks related to the mining process (supply and environmental risks). Additionally, the cathodes of these battery types often require the synthesis of compounds extracted from minerals and well-defined crystal structures. The specific capacity and the operative voltage provided in Table 2 strongly support their influence on the specific energy of the cells. For instance, NMC811, which has the highest specific energy, exhibits excellent properties for both specific capacity and voltage. NMMT and NPBA, expected to have competitive specific energy compared to LFP, also have overlapping specific capacity ranges and similar working voltage to LFP.

Moreover, cathodes and anodes should possess both, high electronic and ionic conductivity. Yet, in Table 2 we observed that contrary to Li-based cathodes such as NCM811, Na-based cathodes such as NVP, NVPF and NMVP suffer generally from poor electronic conductivities[43] (even sometimes ionic conductivities) making it difficult to reach the theoretical capacities at a comparable charging current. Yet, low electronic conductivity is not a result of Na. Analogous cathodes in Li-ion batteries, such as LVP and LVPF, have the same issue. Additionally, material engineering techniques, such as coating, can partially improve electronic conductivity.[44]

In opposite to the electrodes, the electrolyte, which is based on a Li or Na-based salt dissolved in a mixture of organic solvents, should conduct only ions and not electrons between the cathode and anode during the charging and discharging processes. This implies that this electrolyte should have a very low electronic conductivity below $10^{-10}$ S cm$^{-1}$[45] but a very high ionic conductivity typically in the range of $10^{-3}$ to $10^{-2}$ S cm$^{-1}$.[46,47] The current collectors such as Cu or Al with high electronic conductivity, $5.87 \times 10^5$ and $3.55 \times 10^5$ S cm$^{-1}$[48], respectively, facilitates the electron mobility.

The housing of a 21700-cylindrical battery cell is usually metallic, i.e. based on an alloy with Fe and Ni.

Certain chemicals used in the manufacturing of both LIBs and SIBs pose significant health risks due to their toxicity. For instance, N-Methyl-2-pyrrolidone (NMP), used to dissolve the polyvinylidene fluoride (PVDF) binder in electrode manufacturing, is known for its reproductive toxicity and irritating effects, its irritating effects on eyes, skin, and respiratory system.[49] Additionally, hydrofluoric acid (HF), produced from the chemical reactions of lithium hexafluorophosphate (LiPF$_6$) or sodium hexafluorophosphate (NiPF$_6$) with water (which can be present in traces in the electrolyte) or when PVDF catches fire, can be fatal if inhaled or ingested.[49] Furthermore, perfluorinated alkylated substances (PFAS), which may form during complete combustion such as from the recycling process, also exhibit acute toxicity through oral intake and inhalation.[49]

Solid electrolytes for all-solid-state batteries offer a promising solution to improve safety by reducing the probability of leakage and fire risks associated with liquid electrolytes.[50] Various Li-based and Na-based solid electrolytes have been successfully introduced until now as given in Table 3. The solid electrolytes of LIBs including Li$_{1.3}$Al$_{0.3}$Ti$_{1.7}$[PO$_4$]$_3$ (LATP), Li$_{10}$GeP$_2$S$_{12}$ (LGPS), Li$_7$P$_2$S$_8$I, and Li$_{3.25}$Ge$_{0.25}$P$_{0.75}$S$_4$ exhibit good ionic conductivities within the criteria.[46,47] LATP in particular, has a potential stability window between 2.17 – 4.71 V vs. Li$^+$/Li which is comparable to the potential window of lithium salt, LiPF$_6$ of LIBs liquid organic electrolyte. Yet, other high ionic conductivity solid electrolytes such as Li$_7$P$_2$S$_8$I, LPGS and Li$_{3.25}$Ge$_{0.25}$P$_{0.75}$S$_4$ narrower potential windows such as 1.71 – 2.31 V vs. Li$^+$/Li for Li$_7$P$_2$S$_8$I and 1.71 – 2.14 V vs. Li$^+$/Li for LPGS and Li$_{3.25}$Ge$_{0.25}$P$_{0.75}$S$_4$. In contrast, Li$_7$La$_3$Zr$_2$O$_{12}$ (LLZO) offers a lower ionic conductivity than $10^{-3}$ S cm$^{-1}$ but maintains a stable potential window of 0.07 – 3.2 V vs. Li$^+$/Li which is wider than those of Li$_7$P$_2$S$_8$I, LPGS and Li$_{3.25}$Ge$_{0.25}$P$_{0.75}$S$_4$.

.

Table 1 Li and Na-ion battery cells: the materials employed, mass balance, and key performances of 21700-size.

| Properties | NMC811 | LFP | NMMT | NVPF | NMVP | NPBA |
|---|---|---|---|---|---|---|
| Cell mass (g) including cathode, anode, electrolyte, current collector, and housing. | 66.24[a] | 61.64[a] | 56.18[a] | 53.15[a] | 56.00[c] | 44.97[a] |
| Cathode material | LiNi$_{0.8}$Mn$_{0.1}$Co$_{0.1}$O$_2$[a] | LiFePO$_4$[a] | Na$_{1.1}$(Ni$_{0.3}$Mn$_{0.5}$Mg$_{0.05}$Ti$_{0.05}$)O$_2$[a] | Na$_3$V$_2$(PO$_4$)$_2$F$_3$[a] | Na$_4$MnV(PO$_4$)$_3$[b] | Na$_2$FeFe(CN)$_6$[a] |
| Total cathode mass (g) | 22.72[a] | 23.54[a] | 18.59[a] | 19.62[a] | 20.72[c] | 13.71[a] |
| Anode material | Graphite[a] | | | Hard carbon[a] | | |

| Properties | NMC811 | LFP | NMMT | NVPF | NMVP | NPBA |
|---|---|---|---|---|---|---|
| Total anode mass (g) | 13.88[a] | 11.00[a] | 14.15[a] | 11.14[a] | 11.24[c] | 9.18[a] |
| Electrolyte | 1M LiPF$_6$ in DMC:EC 1:1 (vol.) [a] | | | 1M NaPF$_6$ in DMC:EC 1:1 (vol.)[a] | | |
| Total electrolyte mass (g) | 8.91[a] | 9.01[a] | 9.25[a] | 9.35[a] | 9.20[c] | 9.37[a] |
| Current collector (anode/cathode) | Copper/aluminium[a] | | | Aluminium /aluminium[a] | | |
| Current collector mass (g) | 8.66/2.61[a] | 6.77/2.04[a] | 4.82[a] | 3.84[a] | 5.38[c] | 3.55[a] |
| Housing | Iron and nickel[a] | | | Iron and nickel[a] | | |
| Total housing mass (g) | 8.56[a] | 8.56[a] | 8.56[a] | 8.56[a] | 8.56[a] | 8.56[a] |
| Energy density (Wh/kg) | 196.68[b] | 138.81[b] | 114.66[b] | 107.58[b] | 99.97[c] | 105.5[b] |
| Predicted energy density (Wh/kg) | N/A | N/A | 160[c] | 122[c] | N/A | 136[c] |
| Voltage (V) | 3.75[b] | 3.20[b] | 3.20[b] | 3.40[b] | 3.40[b] | 3.20[b] |
| Cycle lifetime | 4000[a] | 7000[a] | 4000[a] | 5000[a] | 7000[b] | 7000[a] |
| Potential toxicity of the cell production process | NMP,PFAS,HF[40,41] | | | NMP, PFAS, HF[40,41] | | |

[a] Mass and components data of NMC811, LFP, MMMT, NPBA are derived from the reference[40]; [b] The energy density is recalculated based on the specific capacity and voltage provided in the reference[41]; [c] Data pertaining to NVPF and cell mass have been adapted from the references[40,41]; [d] Source for data presented the predicted energy density from the reference.[42]

**Table 2** Cathode materials: specific capacity, working potential and ionic and electronic conductivities.

| Properties | NMC811 | LFP | NMMT | NVP | NVPF | NMVP | NPBA |
|---|---|---|---|---|---|---|---|
| Material | LiNi$_{0.8}$Mn$_{0.1}$Co$_{0.1}$O$_2$ | LiFePO$_4$ | Na$_{1.1}$(Ni$_{0.3}$Mn$_{0.5}$Mg$_{0.05}$Ti$_{0.05}$)O$_2$ | Na$_3$V$_2$(PO$_4$)$_3$ | Na$_3$V$_2$(PO$_4$)$_2$F$_3$ | Na$_4$MnV(PO$_4$)$_3$ | Na$_2$Fe$_2$(CN)$_6$ |
| Specific capacity (mAh/g) | 200.0-226.9[51–57] | 130.0-168.0[58–62] | 113.0-150.0[63,64] | 103.5-123.1[65–70] | 107.0-131.5[71–78] | 102.0-109.9[79–83] | 100.0-140.0[84–90] |
| Average/predicted specific capacity (mAh/g) | 209.3[51–57]/- | 152.8[58–62]/- | 124.3[63,64]/156[42] | 114.3[65–70]/- | 119.6[71–78]/128[42] | 106.7[79–83]/- | 122.0[84–90]/160[42] |
| Working potential (V) | 2.4-4.5[51–57] | 2.4-4.2[58–62] | 2.0-4.4[63,64] | 2.0-4.2[65–70] | 2.0-4.5[71–78] | 2.5-3.8[79–83] | 2.0-4.3[84–90] |
| Average/predicted voltage (V) | 3.60[51–57]/- | 3.34[58–62]/- | 3.16[63,64]/3.20[42] | 3.14[65–70]/- | 3.36[71–78]/3.40[42] | 3.16/-[79–83] | 3.04[84–90]/3.40[42] |
| Ionic diffusion (cm$^2$/S) | 1.74×10$^{-11}$ – 2.9×10$^{-11}$[91] | 4.31×10$^{-17}$-1.2×10$^{-12}$[92] | N/A[e] | 4.46×10$^{-13}$-2.87×10$^{-12}$[93] | 2.5×10$^{-11}$-2.2×10$^{-9}$[94] | 2.33×10$^{-15}$-3.02×10$^{-12}$[95] | 10$^{-12}$-10$^{-7}$[96] |
| Electronic conductivity (S·cm$^{-1}$) | 1×10$^{-2}$-2×10$^{-2}$[21] 8×10$^{-6}$[97] | 5×10$^{-8}$[98] | N/A[e] | 8×10$^{-8}$[99] | <2×10$^{-11}$[23] | <2×10$^{-11}$[43] | 4.08 × 10$^{-9}$[100] |

[e] Ionic diffusion and electronic conductivity of this composition, $Na_{1.1}(Ni_{0.3}Mn_{0.5}Mg_{0.05}Ti_{0.05})O_2$ are not found.

**Table 3** Solid electrolytes: potential window and ionic conductivity at 25 °C or room temperature. Comparison with the standard $LiPF_6$ and $NaPF_6$ liquid electrolytes.

| Properties | $LiPF_6$ | $NaPF_6$ | $Li_3OCl$ | LATP | LLZO | LGPS | $Li_7P_2S_8I$ | BASE | NZSP | $Na_3SbS_4$ |
|---|---|---|---|---|---|---|---|---|---|---|
| Material | $LiPF_6$ | $NaPF_6$ | $Li_3OCl$ | $Li_{1.3}Al_{0.3}Ti_{1.7}[PO_4]_3$ | $Li_7La_3Zr_2O_{12}$ | $Li_{10}GeP_2S_2$ | $Li_7P_2S_8I$ | Na-β''-$Al_2O_3$ | $Na_3Zr_2Si_2PO_{12}$ | $Na_3SbS_4$ |
| Potential window (V) | 2.5-3.8[24][f] | 1.0-4.60[f] | 0 - 2.55[101] | 2.17-4.21[102] | 0.07-3.2[103] | 1.71-2.14[102] | 1.71-2.31[102] | Upto 5 V[104] | 1.50-4.0[105] | 0.5-5.0[106] |
| Ionic conductivity (S·cm$^{-1}$) | 8.0x10$^{-3}$[24][f] | 8.5x10$^{-3}$[24][f] | 8.5x10$^{-4}$[26] | 3.0x10$^{-3}$[27] | 3.1x10$^{-4}$[28] | 1.2x10$^{-2}$[29] | 1.6x10$^{-3}$[30] | 1.4x10$^{-2}$[104] | 6.4x10$^{-4}$[107] | 2.1x10$^{-3}$[108] |

[f] 1 M $LiPF_6$ and $NaPF_6$ in solution ethylene carbonate (EC) and diethylene carbonate (DEC) 1:1 from the reference.[24]

## 3. Results and Discussion

### 3.1 Supply and environmental risks of materials

The supply and environmental risks are strongly dependent on the market concentration ($S_i$ and $S_j$), the sourcing country of the materials which influences WGI and EPI indexes, the end-of-life recycle rate (EoL$_{RIR}$), and the ability to be substituted by other alternative materials (Substitution index) as shown in Table 4.

**Table 4** The materials supply and environmental risks, top supplier, and market share.

| Elements | Supply risks | Environmental risks | Global top supplier | Market share (%) | EoL$_{RIR}$ (%) | Substitution index |
|---|---|---|---|---|---|---|
| Antimony (Sb) | 1.77 | 1.79 | China | 56.4 | 28 | 0.94 |
| Aluminium (Al) | 1.18 | 0.57 | Australia | 28.4 | 32 | 0.86 |
| Cobalt (Co) | 2.84 | 1.96 | Congo, D.R. | 62.8 | 22 | 0.98 |
| Copper (Cu) | 0.11 | 0.21 | Chile | 27.7 | 55 | 0.71 |
| Fluorine (F) extracted from fluorspar | 1.11 | 2.25 | China | 55.6 | 1 | 0.91 |
| Germanium (Ge) | 1.74 | 4.58 | China | 83.1 | 2 | 0.94 |
| Iron (Fe) from iron ore | 0.48 | 0.63 | Australia | 36.6 | 31 | 0.95 |
| Lanthanum (La) | 1.83 | 3.33 | China | 68.3 | 1 | 0.97 |
| Lithium (Li) | 1.94 | 2.7 | China | 56.2 | 0 | 0.94 |
| Magnesium (Mg) | 4.1 | 4.82 | China | 90.6 | 13 | 0.94 |
| Manganese (Mn) | 1.16 | 0.81 | South Africa | 29.3 | 9 | 1.00 |
| Natural graphite | 1.79 | 3.11 | China | 66.7 | 3 | 0.98 |
| Nickel (Ni) | 0.52 | 0.79 | China | 33.4 | 16 | 0.92 |
| Phosphorus (P) | 3.18 | 4.42 | China | 78.5 | 0 | 0.98 |

| Elements | Supply risks | Environmental risks | Global top supplier | Market share (%) | EoL$_{RIR}$ (%) | Substitution index |
| --- | --- | --- | --- | --- | --- | --- |
| Silicon (Si) extracted from silica sand | 0.28 | 0.87 | United States | 41.0 | 1 | 0.93 |
| Sodium (Na) | N/A[g] | 0.68 | China | 21.7 | N/A[g] | N/A[g] |
| Titanium (Ti) from titanium metal | 1.54 | 1.87 | China | 42.8 | 1 | 1.00 |
| Vanadium (V) | 1.33 | 2.65 | China | 61.6 | 6 | 0.92 |
| Zirconium (Zr) | 0.74 | 0.84 | Australia | 33.6 | 12 | 0.97 |

[g] Assuming that Na is abundant in any location, it is not considered to be at risk of supply shortages as well as the recycling rate and substitution index are negligible.

Mg exhibits the highest supply and environmental risks due to a highly concentrated market, with China accounting for 90.6% of the global supply. Similarly, P has China as its main producer, contributing 78.5% to the global supply chain. Natural graphite, Co, and V also face high supply and environmental risks, attributed to their major suppliers, China and the Democratic Republic of Congo, who provide more than 60% of the global supply. Conversely, Cu presents the lowest supply and environmental risks due to a less concentrated market share. Despite the market of Cu being more concentrated than that for Na, Cu is the most easily recycled and substituted material among those listed, showing the highest recycling rate. Recycled copper covers 55% of the material demand[33] and has the lowest substitution index; easily be substituted. As for Cu, Al also has high recyclable rate but the SR of Al is high due to the high concentration of the market as shown in Table 4.

### 3.2 Aggregated supply and environmental risks for each battery cell type

Figure 1 displays the supply and environmental risks correlated with the energy density of battery cells using a bubble chart. Although LIBs possess the highest energy densities, they are shown to be less sustainable, with both LFP and NMC811 cathode types presenting higher supply and environmental risks. On the contrary, SIBs have lower energy densities but better supply and environmental risks. The major challenge for SIBs lies indeed in their lower energy density compared to LIBs. Among these, the NMMT battery, which provides the highest energy density of 114.7 Wh/kg[40,41] among the SIBs, is still 17.4% and 41.7% lower than the energy densities of LFP and NMC811, respectively. Additionally, the most sustainable battery, NPBA, with an energy density of 105.5 Wh/kg[40,41], is 24.1% and 46.4% lower than those of LFP and NMC811, respectively.

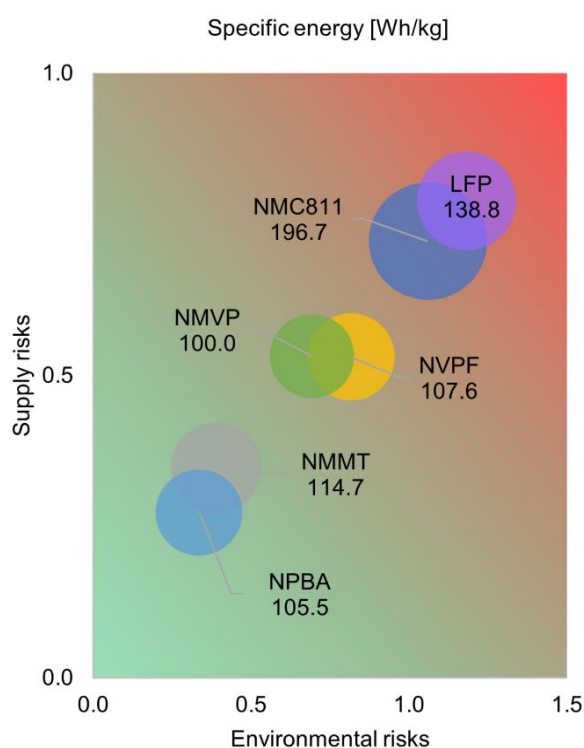

**Figure 1** Bubble chart showing the supply and environmental risks as well as the energy density of battery cells: LiNi$_{0.8}$Mn$_{0.1}$Co$_{0.1}$O$_2$ (NMC811)[40,41], LiFePO$_4$ (LFP)[40,41], Na$_{1.1}$(Ni$_{0.3}$Mn$_{0.5}$Mg$_{0.05}$Ti$_{0.05}$)O$_2$ (NMMT)[40,41], Na$_4$MnV(PO$_4$)$_3$ (NMVP)[40,41], Na$_3$V$_2$(PO$_4$)$_2$F$_3$ (NVPF)[40,41], Na$_2$Fe$_2$(CN)$_6$ (NPBA)[40,41]. The chart uses a gradient color scheme from green in the bottom left to red in the top right, representing the spectrum of sustainability and risk. The green area represents low supply and environmental risks indicating the preferably sustainable materials. In contrast, the red area shows high supply and environmental risks where less sustainable materials are located. The size of the bubbles represents the magnitude of the energy density for each battery cell type. The values are also provided nearby (unit Wh/kg).

Figure 2 illustrates the supply and environmental risks associated with the main components of battery cells, emphasizing that the anode significantly contributes to the higher supply and environmental risks of LIBs compared to SIBs. Li-ion battery anodes are made from graphite, whereas Na-ion battery anodes are produced from hard carbon synthesized from bio-mass eliminating the need for mining. The cathode, another critical component, contributes to supply and environmental risks.

However, these risks are not determined by the type of ion in the cells but rather by the composition of the cathodes proportional to the supply risk and environmental risk per material according to Table 4. It is evident that the cathodes in LFP, NVPF and NMVP-based cells have significantly higher risks compared to other cells because LFP, NVPF, and NMVP contain P, a material associated with high supply and environmental risks. Notably, the NVPF and NMVP cathodes also incorporate V, another material considered to be of high risk. Consequently, these cathode materials possess elevated risks. Conversely, the NMMT and NPBA cathodes present lower risks. In particular, the NPBA cathode relies solely on Fe, a material with significantly lower mining risks.

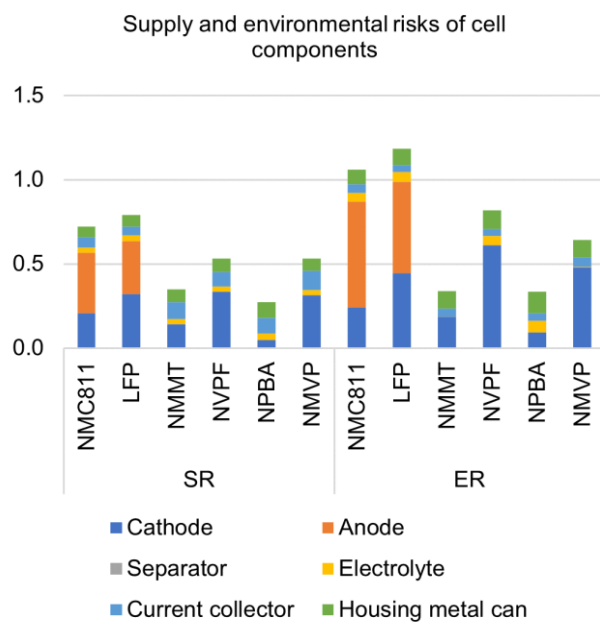

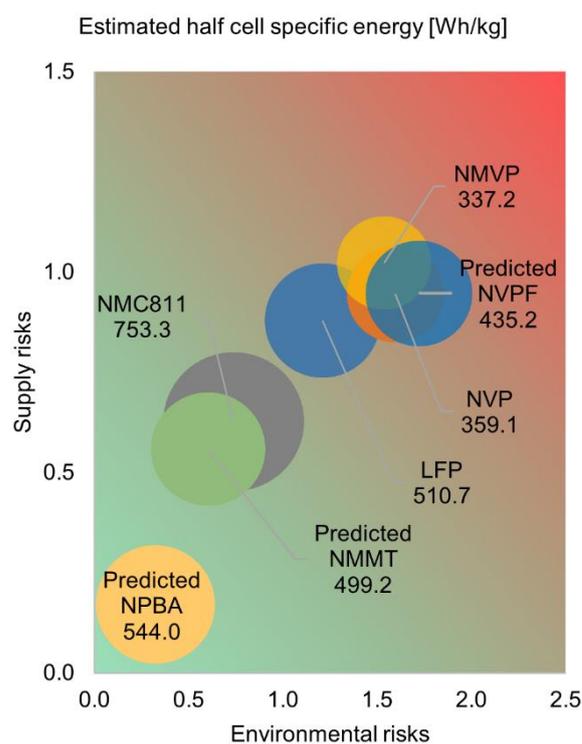

**Figure 2** Supply risks and environmental risks of components of various cell types: $LiNi_{0.8}Mn_{0.1}Co_{0.1}O_2$ (NMC811), $LiFePO_4$ (LFP), $Na_{1.1}(Ni_{0.3}Mn_{0.5}Mg_{0.05}Ti_{0.05})O_2$ (NMMT), $Na_3V_2(PO_4)_2F_3$ (NVPF), $Na_2Fe_2(CN)_6$ (NPBA), $Na_4MnV(PO_4)_3$ (NMVP).

Studies[42,42,63,73,74,78,85,90] show that the specific capacity of Na-ion cathodes can be improved, leading to an improvement in specific energy. In Figure 3, the specific energy of NMC811, LFP, NVP, and NMVP is estimated from the experiments by averaging average specific capacity and average potential. On the other hand, the predicted specific energy of NMMT, NPBA, and NVPF is derived from the predicted specific capacity[42] according to Table 2. The specific energy of the NPBA cathode can surpass the specific energy of LFP. The NPBA cathode exhibits supply and environmental risks at 0.17 and 0.32, respectively. The NPBA risks are approximately three times lower than those of LFP. Similarly, NMMT is another potential Na-ion cathode. Its specific energy is comparable to that of LFP, being only 2.2% lower. The supply and environmental risks of NMMT, at 0.51 and 0.59, are 2.1 and 1.6 times lower than those of LFP.

**Figure 3** Bubble chart showing the supply and environmental risks as well as the estimated half-cell energy density of different cathodes: $LiNi_{0.8}Mn_{0.1}Co_{0.1}O_2$ (NMC811)[51–57], $LiFePO_4$ (LFP)[58–62], $Na_4MnV(PO_4)_3$ (NMVP)[79–83], $Na_3V_2(PO_4)_3$ (NVP)[65–69], predicted of $Na_3V_2(PO_4)_2F_3$ (predicted NVPF)[42], predicted of $Na_{1.1}(Ni_{0.3}Mn_{0.5}Mg_{0.05}Ti_{0.05})O_2$ (predicted NMMT)[42], predicted of $Na_2Fe_2(CN)_6$ (predicted NPBA)[42]. The gradient color scheme from green in the bottom left to red in the top right, represents the spectrum of sustainability and risk. The green area indicates sustainable materials with low supply and environmental risks. The red area shows less sustainable materials with high supply and environmental risks. The size of the bubbles represents the magnitude of the specific energy for each cathode type (unit Wh/kg).

### 3.3 Towards non-toxic and chemically stable solid electrolytes

Developing more sustainable batteries may involve replacing liquid electrolytes with solid electrolytes. Solid electrolytes in general perform better in terms of stability, safety and thermal conductivity. Moreover, they do not leak or evaporate, requiring no extra cooling of the batteries during functioning. Another important advantage is that they allow more versatility in anode material selection reducing the need of carbocenouse materials. Figure 4 summarizes various solid electrolytes used in all-solid-state batteries (ASSBs) technology. Each given solid electrolyte in Figure 4 belongs to one of the three classes established such as halide, sulfides, and oxides.[109] Their sustainability performance as well as their ionic conductivity (a key property of every electrolyte) is compared with the traditional liquid electrolytes $LiPF_6$ and $NaPF_6$ in carbonate solvents. Interestingly, all solid electrolytes are characterized by a lower supply risk as well as lower environmental risks than the traditional liquid electrolyte salts $LiPF_6$ and $NaPF_6$. This is because these solid electrolytes contain either lower-risk materials or require a lower mass fraction of high-risk materials (such as P) compared to liquid electrolytes. For example, the P required in NZSP is only of about 5.8% of its mass, which is much lower than the demand in $LiPF_6$ and $NaPF_6$ at 20.4% and 18.4%, respectively. Moreover, the risks associated with LLZO, the solid electrolyte with the highest supply and environmental risk, mainly stem from La. When comparing the sustainability parameters among the solid electrolytes, those for SIBs pose lower supply and environmental risks than those for LIBs. Their ionic conductivity ranges from 0.6 to 14 mS/cm, which is comparable to those of LIBs. Yet, the highest ionic conductivity is observed for the sodium-β-alumina (BASE) electrolyte which surpasses the ionic conductivity of LGPS used in Li-based ASSBs. Sadly, except for LGPS and BASE, other solid electrolytes have a much smaller ionic conductivity than the traditional liquid electrolytes $LiPF_6$ and $NaPF_6$. This is mainly due to Na-based secondary phase formation[110] or strong Li accumulation at the crystallographic defects (such as grain boundaries).[111] This reduces strongly the Li or Na mobility and conductivity within the solid electrolyte as proved recently by our studies done with atom probe tomography and electrochemical impedance spectroscopy.[111] Hence, the task in the next years is to improve the ionic conductivity of such solid electrolytes to achieve not only highly sustainable but also highly performant solid electrolytes; an important requirement for a successful implementation of ASSBs in e-mobility applications.

and environmental risk, respectively, making the chart more intuitive to read. Cells with better properties are closer to the outer circle. The NMC811 cell exhibits the highest energy density among all cells but is less sustainable than Na-ion battery cells, as indicated by its lower cycle lifetime, supply stability, and environmental compatibility. In addition, two Na-ion battery cells, NMMT and NPBA, stand out for their high supply stability and environmental compatibility. NMMT boasts a higher energy density and voltage compared to NPBA, but its downside is a lower cycle lifetime. Yet, one needs to note here that layered oxide cathodes have not reached their technology maturity and that their performance is expected to improve in the future. Moreover, there is potential for enhancing the energy densities of NMMT and NPBA in the future, potentially surpassing that of LFP as illustrated in Figure 5 making these cell types more appealing for commercialization.

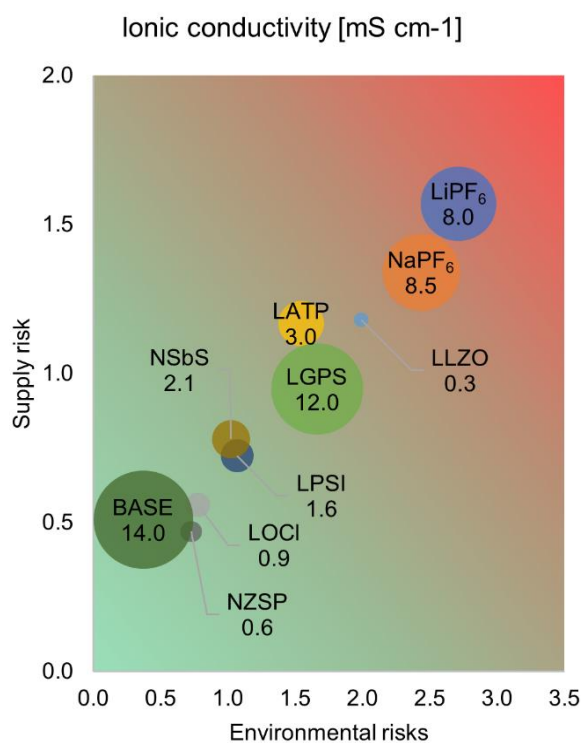

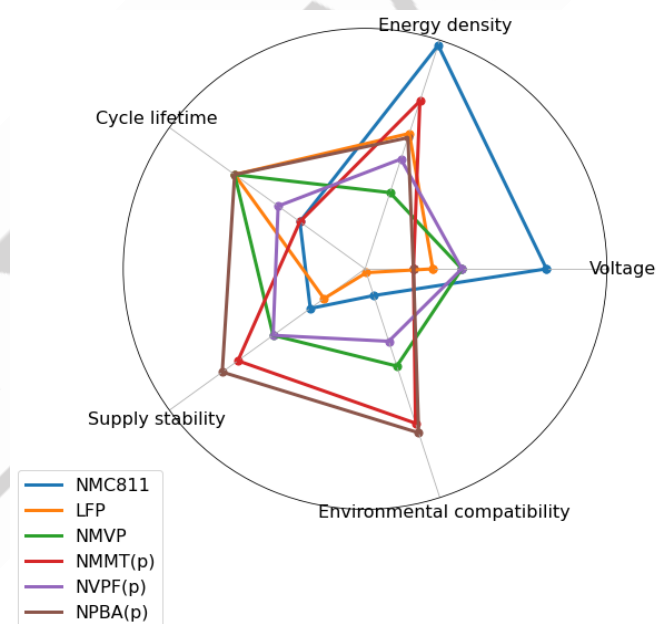

**Figure 5** Radar chart compares between performance and sustainability of LIB and SIB cells. The energy densities of $LiNi_{0.8}Mn_{0.1}Co_{0.1}O_2$ (NMC811)[40,41], $LiFePO_4$ (LFP)[40,41], $Na_4MnV(PO_4)_3$ (NMVP)[40,41], predicted of $Na_{1.1}(Ni_{0.3}Mn_{0.5}Mg_{0.05}Ti_{0.05})O_2$ (NMMT(p))[42], predicted of $Na_3V_2(PO_4)_2F_3$ (NVPF(p))[42], and predicted of $Na_2Fe_2(CN)_6$ (NPBA(p))[42] are represented.

**Figure 4** Bubble chart showing the ionic conductivity, supply, and environmental risks of liquid electrolytes $LiPF_6$[24] and $NaPF_6$[25] as well as of solid electrolytes for Li-based ASSBs including $Li_3OCl$ (LOCl)[26], $Li_{1.3}Al_{0.3}Ti_{1.7}[PO_4]_3$ (LATP)[27], $Li_7La_3Zr_2O_{12}$ (LLZO)[28], $Li_{10}GeP_2S_{12}$ (LGPS)[29], $Li_7P_2S_8I$ (LPSI)[30], and for the Na-based ASSBs including Na-β"-$Al_2O_3$ (BASE)[104], $Na_3Zr_2Si_2PO_{12}$ (NZSP)[107], and $Na_3SbS_4$ (NSbS)[108]. The chart uses a gradient color scheme from green in the bottom left to red in the top right, representing the spectrum of sustainability and risk. The green area shows sustainable materials with low supply and environmental risks, while the red area represents less sustainable materials with high supply and environmental risks. the size of the bubble corresponds to the magnitude of the ionic conductivity (unit mS/cm).

### 3.4 Correlation between performance and sustainability for LIB and SIB cells

In Figure 5, the performance and sustainability of Li-ion and Na-ion battery cells are compared. The metrics of supply stability and environmental compatibility are derived inversely from supply risk

## Conclusion

Material supply risk is associated with factors such as market concentration, geographical sources of materials, reliance on imports, substitutability with alternative materials, and recyclability. A material with lower recyclability forces the market to rely more on primary raw materials, increasing supply risk. Additionally, if materials must be imported from markets with high concentration, supply risks escalate due to fewer alternative sources in the event of a shortage from main suppliers. Environmental risk, similarly, is influenced by the volume of primary raw materials used and is exacerbated when these materials are sourced from countries with poor environmental management.

Na-ion battery cells outperform Li-ion cells in terms of sustainability, primarily because they require fewer minerals, especially at the anode, eliminating the need for graphite, which poses a significant supply risk. However, some Na-ion battery cells, such as NMVP and NVPF, have less sustainable cathodes due to the high content of materials at high risk, like P and V. When considering the cathodes in isolation, their impact on sustainability is negatively significant. Conversely, Na-ion battery cells like NMMT and NPBA-based ones, coupled with an HC anode, demonstrate substantial sustainable potential, though their current drawbacks include low energy density and, for NMMT, reduced cycle lifetime. Nonetheless, it is anticipated that improvements in their performance—specifically, energy density—will eventually surpass that of LFP cells.

## Acknowledgements

The authors acknowledge the SSE (Sustainable System Engineering) program of the University of Freiburg for providing the financial support to conduct these studies.

**Keywords:** material criticality • supply risk • environmental risk • sodium-ion batteries • lithium-ion batteries•

13